\documentclass[conference]{IEEEtran}
\IEEEoverridecommandlockouts
\usepackage{cite}
\usepackage{amsmath,amssymb,amsfonts}
\usepackage{algorithmic}
\usepackage{graphicx}
\usepackage{textcomp}
\usepackage{xcolor}
\def\BibTeX{{\rm B\kern-.05em{\sc i\kern-.025em b}\kern-.08em
    T\kern-.1667em\lower.7ex\hbox{E}\kern-.125emX}}

\begin{document}

\title{Throughput Optimization of Intelligent Reflecting Surface Assisted User Cooperation in WPCNs }
\author{Yuan~Zheng\IEEEauthorrefmark{1},~Suzhi~Bi\IEEEauthorrefmark{1}\IEEEauthorrefmark{2},~Ying-Jun~Angela~Zhang\IEEEauthorrefmark{3}, and Hui~Wang\IEEEauthorrefmark{1}\\
\IEEEauthorrefmark{1}College of Electronic and Information Engineering, Shenzhen University, Shenzhen, 518060, China \\
 \IEEEauthorrefmark{2}Peng Cheng Laboratory, Shenzhen, 518066, China \\
\IEEEauthorrefmark{3}Department of Information Engineering, The Chinese University of Hong Kong, Shatin, N.T., Hong Kong SAR\\
 E-mail:~\IEEEauthorrefmark{1}\{zhyu, bsz, wanghsz\}@szu.edu.cn, \IEEEauthorrefmark{3}yjzhang@ie.cuhk.edu.hk
%\thanks{This work was supported in part by the National Key Research and Development Program (Project 2019YFB1803305), the National Natural Science Foundation of China (Project 61871271), the General Research Funding (project number 14200315) established by Hong Kong Research Grant Council, the Guangdong Province Pearl River Scholar Funding Scheme 2018 (Project 308/00003704), the Foundation of Shenzhen City (Project JCYJ20170818101824392, JCYJ20190808120415286), and the Science and Technology Innovation Commission of Shenzhen (Project 827/000212).}
}

\maketitle

\begin{abstract}
Intelligent reflecting surface (IRS) can effectively enhance the energy and spectral efficiency of wireless communication system through the use of a large number of low-cost passive reflecting elements. In this paper, we investigate throughput optimization of IRS-assisted user cooperation in a wireless powered communication network (WPCN), where the two WDs harvest wireless energy and transmit information to a common hybrid access point (HAP). In particular, the two WDs first exchange their independent information with each other and then form a virtual antenna array to transmit jointly to the HAP. We aim to maximize the common (minimum) throughput performance by jointly optimizing the transmit time and power allocations of the two WDs on wireless energy and information transmissions and the passive array coefficients on reflecting the wireless energy and information signals. By comparing with some existing benchmark schemes, our results show that the proposed IRS-assisted user cooperation method can effectively improve the throughput performance of cooperative transmission in WPCNs.

\end{abstract}

\section{Introduction}
To meet the increasing device energy consumption of modern wireless networks,  wireless powered communication network (WPCN) has been recently proposed and widely studied (e.g.,\cite{2014:Bi,2014:Ju1,2018:Bi,2019:Bi,2018:Y}), which uses dedicated wireless energy transferring nodes to power the operation of communication devices. Compared to its conventional battery-powered counterpart, WPCN has shown its advantages in lowering the network operating cost and improving the robustness of communication service especially in low power applications, such as sensor and internet of things (IoT) networks. The major technical challenge in WPCNs is the low power transfer efficiency over long distance, which also leads to severe user unfairness problem in the achievable data rates due to the dissimilar user locations. To tackle the user unfairness problem, many user cooperation methods have been proposed and have demonstrated their effectiveness in varies network setups \cite{2017:Zhong,2014:Ju2,2019:Zheng,2020:Yuan}. Nonetheless, the low energy transfer efficiency is still a fundamental performance bottleneck of WPCN systems.

Recently, with the developments in meta-surface technology \cite{2014:Tj}, intelligent reflecting surface (IRS) technology has received widespread attention in wireless communications\cite{2019:WQQ1}. In particular, an IRS comprises a massive number of  reconfigurable reflecting elements and a smart controller, where each element can reflect impinging electromagnetic waves with a controllable phase shift using the IRS controller. By properly adjusting the phase shifts of the elements of IRS, the reflected signals can be coherently combined with those from other paths at the receiver to maximize the signal strength. Combing the virtual array gain and the reflect beamforming gain, the IRS is capable of enhancing wireless energy transfer efficiency and therefore fulfilling the potential of WPCNs.

 Previous studies have reported the application of the IRS in wireless communications\cite{2018:WQQ2,2019:Huang,2019:WQQ,2019:QUA}. For instance, \cite{2018:WQQ2} considered a joint design of active beamforming at the base station (BS) and passive beamforming at the IRS to minimize the total transmit power. \cite{2019:Huang} studied the transmit power allocation and passive beamforming design to maximize energy/spectral efficiency. \cite{2019:WQQ} proposed to use a set of distributed IRSs to assist simultaneous wireless information and power transfer (SWIPT) from a multi-antenna AP to multiple information receivers and energy receivers. Multiple-input multiple-output (MIMO) beamforming was investigated in \cite{2019:QUA} for IRS-assisted systems, where the phase shifts were either given or designed only for rank-one BS-to-IRS line-of-sight (LOS) channel. However, most of the existing works only exploit IRS for enhancing the received signal strength in the forward links (FLs) from the BS to the users. In practice, the reflection of IRS is also applicable to improve the spectral efficiency in the reverse links (RLs).

In this paper, we investigate a novel IRS-assisted two-user cooperation method in WPCNs. As shown in Fig.~\ref{Fig.1}, we consider that an HAP broadcasts wireless energy to two WDs in the FLs and receives cooperative information transmission in the RLs. Specifically, the two WDs first exchange their independent information with each other and then form a virtual antenna array to transmit jointly to the HAP. In this case, the IRS is used to assist the wireless energy transfer (WET), the information exchange among the two WDs, and the joint wireless information transmission (WIT) to the HAP. With the proposed IRS-assisted cooperation method, we formulate an optimization problem to maximize the common throughput of the two users, which is an important metric of user fairness in WPCN. This involves a joint optimization of the transmit time, power allocation of the two WDs on wireless energy and information transmissions, and the passive array coefficients to reflect the wireless energy and information signals. We propose an efficient method to tackle the non-convex problem. Besides, we conduct extensive simulations to show that the proposed IRS-assisted method can effectively enhance the throughput performance in WPCN compared with some representative benchmark methods.

\section{System Model}
\subsection{Channel Model}
\begin{figure}
  \centering
   \begin{center}
      \includegraphics[width=0.45\textwidth]{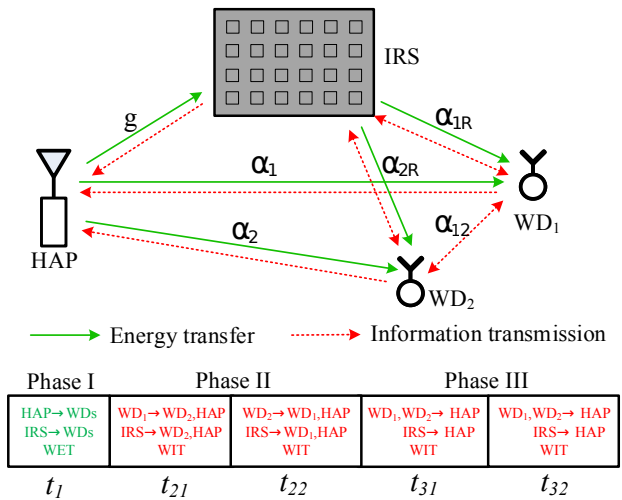}
   \end{center}%\vspace{-.2in}
  \caption{The network structure and transmit protocol of the proposed IRS-assisted cooperation scheme.}
  \label{Fig.1}
\end{figure}
As show in Fig.~\ref{Fig.1}, we consider a WPCN consisting of one HAP and two WDs denoted by WD$_1$ and WD$_2$, where the HAP and WDs have a single antenna each. Both WDs first harvest RF energy in the FL and then transmit information in the RL. To enhance the propagation performance, we employ an IRS composed of $N$ passive elements in the vicinity of the devices to assist the transmissions of the WPCN. The IRS can dynamically adjust the phase shift of each reflecting element based on the propagation environment learned through periodic sensing. Due to the substantial path loss, we only consider one-time signal reflection by the IRS and ignore the signals that are reflected thereafter.

For simplicity, we assume that all channels experience quasi-static flat fading and follow channel reciprocity between the FL and RL. The baseband equivalent channels from the HAP to IRS, from IRS to WD$_i$, from the HAP to WD$_i$ and from WD$_1$ to WD$_2$ are denoted by $\mathbf g\in \mathbb {C}^{1\times N}$, ${\boldsymbol\alpha_{iR}}\in \mathbb {C}^{N\times 1}, i=1,2$, $\alpha_i$ and $\alpha_{12}$, respectively. We denote $g=\|\mathbf g\|^2$, ${ h_{iR}}=\|{{\boldsymbol\alpha_{iR}}}\|^2$,  $h_{i} = |\alpha_{i}|^2$ and $h_{12} = |\alpha_{12}|^2$ as the corresponding channel gains. It is assumed that the channels of different transceiver pairs are independent to each other. Besides, the entries inside all channel vectors are modeled as zero-mean independent and identically distributed (i.i.d.) complex Gaussian random variables with variance depending on the path loss of the respective wireless links.
We denote $\boldsymbol\Theta=\rm{diag}$$(\beta_1v_1,\cdots,\beta_n v_n,\cdots,\beta_N v_N)$ with $v_n=e^{j\theta_n},n=1,\cdots,N$ as the diagonal reflection coefficient matrix at the IRS, where $\beta_n \in [0, 1]$ and $\theta_n \in [0, 2\pi]$ are the amplitude coefficient and phase shift of each element, respectively (diag($\mathbf a$) denotes a diagonal matrix with its diagonal elements given in the vector $\mathbf a$).  We assume that both the HAP and IRS have perfect channel state information (CSI) like in \cite{2018:WQQ2,2019:Huang,2019:WQQ,2019:QUA,2020:Zheng}.

\subsection{Protocol Description}

As shown in Fig.~\ref{Fig.1}, we consider a three-phase operating protocol of the proposed scheme.  In the first phase of duration $t_1$, the HAP transfers wireless energy in the FL for the two WDs to harvest.  Meanwhile, the IRS scatters the incident signal from the HAP to the WDs, such that the WDs receive both the direct-path and reflect-path signals from the HAP. In the second phase of duration $t_{21}$ and $t_{22}$,  WD$_1$ and WD$_2$ exchange their information to each other with the help of IRS. In the last phase of length $t_3$, WD$_1$ and WD$_2$ jointly transmit their information to the HAP with the help of IRS. Specifically, the two WDs first jointly transmit the information of WD$_1$ and then the information of WD$_2$ with duration $t_{31}$ and $t_{32}$, respectively, with $t_3=t_{31}+t_{32}$. Note that we have a total time constraint
\begin{equation}
\label{t}
\small
 {t_1} + {t_{21}} + t_{22} + {t_{31}}  +{t_{32}}  \le T.
\end{equation}

As no inter-user interference exists in the above mentioned transmission scheme, we set $\beta_n=1$  to maximize the signal reflection by the IRS, i.e., $\boldsymbol\Theta=\rm{diag}$$(v_1,\cdots,v_n,\cdots, v_N)$ with $|v_n|=1,n=1,\cdots,N$. In the following section, we derive the optimal throughput performance of the considered IRS-assisted user cooperation in WPCN.

\section{Throughput Performance Analysis}
\subsection{Phase I: Energy Transfer}
In the first stage of length $t_1$, the HAP transmits energy with fixed power $P_1$ in $t_1$ amount of time. We denote the energy
signal as $x_0(t)$ with $E[|x_0(t)|^2] = 1$. The signal received at WD$_i$ can be expressed as \cite{2018:WQQ2}
\begin{equation}
y_i^{(1)}(t) = (\mathbf g\boldsymbol\Theta_1{\boldsymbol\alpha_{iR}} +\alpha_i)\sqrt {P_1}x_0(t)+n_i^{(1)}(t), i=1,2,
\end{equation}
where $\boldsymbol\Theta_1=\text{diag} (v_{1,1},\cdots,v_{1,N})$ denotes the energy reflection coefficient matrix at the IRS with $|v_{1,n}|=1$ for $n=1,\cdots, N$, $n_i^{(1)}(t)$ denotes the receiver noise. The amount of energy harvested by the $i$-th WD is given by
\begin{equation}
\label{energy}
\begin{aligned}
E_i={\eta}E[|y_i^{(1)}(t)|^2]{t_1}=\eta |\mathbf g\boldsymbol\Theta_1{\boldsymbol\alpha_{iR}}+\alpha_i|^2P_1 t_1, i=1,2,
\end{aligned}
\end{equation}
where $0\textless \eta \textless 1$ denotes the fixed energy harvesting efficiency.\footnote{Although a single energy harvesting circuit exhibits non-linear energy harvesting property due to the saturation effect of circuit, it is shown that the non-linear effect can be effectively rectified by using multiple energy harvesting circuits concatenated in parallel, resulting in a sufficiently large linear conversion region in practice \cite{2019:Kang}. }

\subsection{Phase II: Information Exchange}
During the information exchange phase, WD$_1$ and WD$_2$ transmit their information to each other with the transmit power $P_{21}$ and $P_{22}$ for $t_{21}$ and $t_{22}$ amount of time, respectively. We denote $x_1(t)$ as the transmitted signal from WD$_1$ with $ E[|x_1(t)|^2]= 1$.  Then,  the signal received at WD$_2$ and the HAP are expressed as, respectively,
\begin{equation}
\label{y22}
y_2^{(2)}(t)=({\boldsymbol\alpha_{2R}^T} \boldsymbol\Theta_2{\boldsymbol\alpha_{1R}}+\alpha_{12})\sqrt{P_{21}}{ x_1(t)}+n_2^{(2)}(t),
\end{equation}
\begin{equation}
y_0^{(21)}(t)=(\mathbf g\boldsymbol\Theta_2{\boldsymbol\alpha_{1R}}+\alpha_1)\sqrt{P_{21}}{ x_1(t)}+ n_0^{(2)}(t),
\end{equation}
where $\boldsymbol\Theta_2=\text{diag} (v_{2,1},\cdots,v_{2,N})$ denotes the reflection-coefficient matrix at the IRS in duration $t_{21}$ with $|v_{2,n}|=1,n=1,\cdots,N$, $n_2^{(2)}(t)$ and $n_0^{(2)}(t)$ denote the receiver noise with power $N_0$, $(\cdot)^T$ denotes the transpose operator. Then, the achievable rates from WD$_1$ to WD$_2$ and the HAP are
\begin{equation}
\label{R12}
R_1^{(2)} = \frac{t_{21}}{T}{\log _2}\left(1 + \frac{P_{21}|{\boldsymbol\alpha_{2R}^T} \boldsymbol\Theta_2{\boldsymbol\alpha_{1R}}+\alpha_{12}|^2}{N_0}\right),\\
\end{equation}
\begin{equation}
\label{R021}
R_0^{(21)} =\frac {t_{21}}{T}{\log _2}\left(1 + \frac{P_{21}|\mathbf g\boldsymbol\Theta_2{\boldsymbol\alpha_{1R}}+\alpha_1|^2}{N_0}\right).
\end{equation}

Similarly, let $\boldsymbol\Theta_3= \text{diag} (v_{3,1},\cdots,v_{3,N})$ denote the reflection-coefficient matrix at the IRS when WD$_2$ transmits with duration $t_{22}$, where $|v_{3,n}|=1,n=1,\cdots,N$.
 Then, the achievable rates from WD$_2$ to WD$_1$ and the HAP are
\begin{equation}
\label{R22}
R_2^{(2)} = \frac{t_{22}}{T}{\log_2}\left(1 + \frac{P_{22}|\boldsymbol\alpha_{1R}^T\boldsymbol\Theta_3{\boldsymbol\alpha_{2R}}+\alpha_{12}|^2}{N_0}\right),
\end{equation}
\begin{equation}
\label{R022}
R_0^{(22)} = \frac{t_{22}}{T}{\log_2}\left(1 + \frac{P_{22}|\mathbf g\boldsymbol\Theta_3{\boldsymbol\alpha_{2R}}+\alpha_2|^2}{N_0}\right).
\end{equation}
\subsection{Phase III: Joint Information Transmission}
In the last phase of duration $t_3$, the two WDs jointly transmit their information to the HAP. Meanwhile, the IRS reflects signal of the two WDs to the HAP. Specifically, WD$_i$ transmits with power $P_{3i}$ for $t_{3i}$ amount of time for $i=1,2$. Thus, the total energy consumption of WD$_i$ in Phase II and III is restricted by
\begin{equation}
\label{con}
\begin{aligned}
t_{2i}P_{2i} + (t_{31}+t_{32})P_{3i}\le E_i, i=1,2.
\end{aligned}
\end{equation}

 In this stage, we consider Alamouti space-time block code transmit diversity scheme for joint information transmission, where the achievable rates from WD$_1$ and WD$_2$ to the HAP are
\begin{equation}
\begin{aligned}
\label{R13}
R_1^{(3)} =\frac{t_{31}}{T}{\log _2}\bigg(1 &+ \frac{P_{31}|\mathbf g\boldsymbol\Theta_4{\boldsymbol\alpha_{1R}}+\alpha_1|^2}{N_0}\\&+\frac{P_{32}|\mathbf g\boldsymbol\Theta_4{\boldsymbol\alpha_{2R}}+\alpha_2|^2}{N_0}\bigg),
\end{aligned}
\end{equation}
\begin{equation}
\begin{aligned}
\label{R23}
R_2^{(3)} =\frac{t_{32}}{T}{\log _2}\bigg(1 &+ \frac{P_{31}|\mathbf g\boldsymbol\Theta_4{\boldsymbol\alpha_{1R}}+\alpha_1|^2}{N_0}\\&+\frac{P_{32}|\mathbf g\boldsymbol\Theta_4{\boldsymbol\alpha_{2R}}+\alpha_2|^2}{N_0}\bigg),
\end{aligned}
\end{equation}
where $\boldsymbol\Theta_4= \text{diag} (v_{4,1},\cdots,v_{4,N})$ denotes the reflection-coefficient matrix at the IRS with $|v_{4,n}|=1,n=1,\cdots,N$.

Overall, the achievable rate of WD$_i$ is \cite{2014:Ju2}
\begin{equation}
\label{11}
\begin{aligned}
 R_i = \min \ \left[R_i^{(2)}, R_0^{(2i)}+ R_i^{(3)}\right],i=1,2.
\end{aligned}
\end{equation}
Without loss of generality, we assume $T=1$ in this paper.

\section{Common Throughput Maximization}
\subsection{Problem Formulation}
In this section, we focus on maximizing the common (minimum) throughput of the two WDs by jointly optimizing the transmit time allocation $\mathbf{t}=[t_1,t_{21},t_{22},t_{31},t_{32}]$, power allocation $\mathbf{P}=[P_{21},P_{22},P_{31},P_{32}]$ and the phase shift matrices $\widetilde{\boldsymbol\Theta}=[\boldsymbol\Theta_1,\boldsymbol\Theta_2, \boldsymbol\Theta_3,\boldsymbol\Theta_4]$, i.e.,
\begin{equation}
\begin{aligned}
\label{1}
   ~~~~~~(\rm{P1}): & \max_{\mathbf{t,P},\widetilde{\boldsymbol\Theta}} & & \min \ (R_1,R_2)\\
    &~~\text{s. t.}  & &(\ref{t}) , (\ref{energy}) \ \text{and}\ (\ref{con}),\\
    &  & & t_1,t_{2i},t_{3i},P_{2i},P_{3i}\geq 0,i=1,2,\\
    & & & |v_{i,n}|=1, i=1,2,3,4, n=1,\cdots,N.~~~~~~~~
\end{aligned}
\end{equation}
 By introducing an auxiliary variable $\overline{R}$, (P1) can be equivalently rewritten as
\begin{equation}
\begin{aligned}
 ~~(\rm{P2}): &~~~\max_{\overline{R},\mathbf{t,P},\widetilde{\boldsymbol\Theta}} & &  \overline{R}\\
    &~~~~~~\text{s. t.}     & &(\ref{t}) , (\ref{energy}) \ \text{and}\ (\ref{con}),\\
  & & & t_1,t_{2i},t_{3i},P_{2i},P_{3i}\geq 0,i=1,2,\\
  & & & \overline{R}\leq R_1^{(2)},\ \overline{R}\leq  R_0^{(21)}+R_1^{(3)},\\
 & & & \overline{R}\leq R_2^{(2)}, \ \overline{R}\leq  R_0^{(22)}+R_2^{(3)},\\
  & & &|{v}_{i,n}|=1,\;i=1,2,3,4,n=1,\cdots,N.~~~~~
\end{aligned}
\end{equation}
Notice that all the achievable data rates expressions of WD$_1$ and WD$_2$ are not concave functions. Besides, (\ref{energy}), (\ref{con}) and the modulus constraint of ${v}_{i,n}$ are also not convex. Therefore, (P2) is a non-convex problem in its current form, which lacks of efficient optimal algorithms. In the next subsection, we transform the above non-convex problem into a convex one using semidefinite relaxation technique.
\subsection{Proposed Solution to (P2)}
To solve the non-convex problem (P2), let $\boldsymbol v_i=[v_{i,1},\cdots,v_{i,N}], i=1,2,3,4$. Then, we have $\mathbf g\boldsymbol\Theta_i{\boldsymbol\alpha_{jR}}={\boldsymbol {v}_{i}}\text{diag}(\mathbf g){\boldsymbol\alpha_{jR}},j=1,2$, ${\boldsymbol\alpha_{2R}^T }\boldsymbol\Theta_2{\boldsymbol\alpha_{1R}}={\boldsymbol {v}_{2}}\text{diag}({\boldsymbol\alpha_{2R}^T}){\boldsymbol\alpha_{1R}}$ and ${\boldsymbol\alpha_{1R}^T }\boldsymbol\Theta_2{\boldsymbol\alpha_{2R}}={\boldsymbol {v}_{2}}\text{diag}({\boldsymbol\alpha_{1R}^T}){\boldsymbol\alpha_{2R}}$. To tackle the non-convex modulus constraint in (P2),  we first define $\bar{\boldsymbol v}_i=\left[\begin{matrix}\boldsymbol v_i^T\\1\end{matrix}\right],i=1,2,3,4$, $\bar{\boldsymbol \gamma}_j=\left[\begin{matrix}\boldsymbol \gamma_j\\ \alpha_{j}\end{matrix}\right],j=1,2$, $\bar{\boldsymbol \gamma}_{2j}=\left[\begin{matrix}\boldsymbol \gamma_{2j}\\ \alpha_{12}\end{matrix}\right]$, $\mathbf V_i=\bar{\boldsymbol v}_i \bar{\boldsymbol v}_i^H$, $\boldsymbol \psi_{j}=\bar{\boldsymbol \gamma}_{j}\bar{\boldsymbol \gamma}_{j}^H$ and $\boldsymbol \psi_{2j}=\bar{\boldsymbol \gamma}_{2j}\bar{\boldsymbol \gamma}_{2j}^H$, where $(\cdot)^H$ denotes the complex conjugate operator.  Thus, we have
 \begin{equation}
 \begin{aligned}
  |\mathbf g\boldsymbol\Theta_i{\boldsymbol\alpha_{jR}} +\alpha_j|^2=|{\boldsymbol v}_i \boldsymbol \gamma_{j}&+\alpha_j|^2=\text{tr}(\mathbf V_i\boldsymbol \psi_{j}),\\&i=1,2,3,4,j=1,2,
\end{aligned}
 \end{equation}
 \begin{equation}
 \label{r2}
   |{\boldsymbol\alpha_{2R}^T }\boldsymbol\Theta_2{\boldsymbol\alpha_{1R}}+\alpha_{12}|^2=|{\boldsymbol v}_2 \boldsymbol \gamma_{21}+\alpha_{12}|^2=\text{tr}(\mathbf V_2\boldsymbol \psi_{21}),
 \end{equation}
  \begin{equation}
 \label{r2}
   |{\boldsymbol\alpha_{1R}^T }\boldsymbol\Theta_3{\boldsymbol\alpha_{2R}}+\alpha_{12}|^2=|{\boldsymbol v}_3 \boldsymbol \gamma_{22}+\alpha_{12}|^2=\text{tr}(\mathbf V_3\boldsymbol \psi_{22}),
 \end{equation}
where $\boldsymbol \gamma_{21}=\text{diag}({\boldsymbol\alpha_{2R}^T}){\boldsymbol\alpha_{1R}}$, $\boldsymbol \gamma_{22}=\text{diag}({\boldsymbol\alpha_{1R}^T}){\boldsymbol\alpha_{2R}}$ and $\boldsymbol \gamma_{j}=\text{diag}(\mathbf g){\boldsymbol\alpha_{jR}},j=1,2$.

 Next, we introduce auxiliary  variables $\tau_{2j}=t_{2j}P_{2j}, \tau_{3j}=t_{3j}P_{3j}, j=1,2$, $\tau_{31}'=t_{32}P_{31}$ and $\tau_{32}'=t_{31}P_{32}$.
Note that $\left[\mathbf V_i\right]_{n,n}=1, i=1,2,3,4, n=1,\cdots,N+1$ hold from the modulus constraint of $v_{i,n}$ ($\left[\mathbf X\right]_{m,n}$ denotes the element in the $m$-th row and $n$-th column of matrix $\mathbf X$).
 Then, we define $\mathbf W_1=t_1\mathbf V_1\succeq 0$, $\mathbf W_{2}=\tau_{21}\mathbf V_2\succeq 0$, $\mathbf W_{3}=\tau_{22}\mathbf V_3\succeq 0$, $\mathbf W_{4j}=\tau_{3j}\mathbf V_4\succeq 0, \mathbf W_{4j}'=\tau_{3j}'\mathbf V_4\succeq 0, j=1,2$, which satisfy the following constraints
\begin{equation}
\begin{aligned}
\label{u}
&[\mathbf W_1]_{n,n}=t_1,n=1,\cdots,N+1,\\
&[\mathbf W_{2}]_{n,n}=\tau_{21},[\mathbf W_{3}]_{n,n}=\tau_{22},\\
&[\mathbf W_{4j}]_{n,n}=\tau_{3j}, [\mathbf W_{4j}']_{n,n}=\tau_{3j}',j=1,2.
\end{aligned}
\end{equation}
Therefore, we can re-express $R_1^{(2)}$ in (\ref{R12}) as
\begin{equation}
\begin{aligned}
\label{Ry11u}
   R_1^{(2)} &= t_{21}\log_{2}\left(1+\rho\frac{|{\boldsymbol v}_2 \boldsymbol \gamma_{21}+\alpha_{12}|^2\tau_{21}}{t_{21}}\right),\\
   &=t_{21}\log_{2}\left(1+{\rho}{\frac{\text{tr}\left(\boldsymbol \psi_{21}\mathbf V_{2}\right)\tau_{21}}{t_{21}}}\right),\\
   &=t_{21}\log_{2}\left(1+{\rho}{\frac{\text{tr}\left(\boldsymbol \psi_{21}\mathbf W_{2}\right)}{t_{21}}}\right),
\end{aligned}
 \end{equation}
where $\rho=\frac{1}{N_0}$ is a constant. Similarly, the achievable data rates in (\ref{R021})-(\ref{R022}), (\ref{R13}) and (\ref{R23}) can be equivalently reformed as
\begin{equation}
  \label{Ry021u}
   R_0^{(21)} = t_{21}\log_{2}\left(1+{\rho}{\frac{\text{tr}\left(\boldsymbol \psi_{1}\mathbf W_{2}\right)}{t_{21}}}\right),
 \end{equation}
\begin{equation}
\label{Ry22V}
R_2^{(2)}=t_{22}\log_{2}\left(1+\rho\frac{\text{tr}\left(\boldsymbol \psi_{22}\mathbf W_{3}\right)}{t_{22}}\right),
\end{equation}
 \begin{equation}
\label{Ry022u}
   R_0^{(22)} = t_{22}\log_{2}\left(1+{\rho}{\frac{\text{tr}\left(\boldsymbol \psi_{2}\mathbf W_{3}\right)}{t_{22}}}\right),
 \end{equation}
 \begin{equation}
\label{Ry144u}
R_1^{(3)}={t_{31}}\log_{2}\left(1+\rho\frac{\text{tr}\left(\boldsymbol \psi_{1}\mathbf W_{41}\right)}{t_{31}}+\rho\frac{\text{tr}\left(\boldsymbol \psi_{2}\mathbf W_{42}'\right)}{t_{31}}\right),
\end{equation}
\begin{equation}
\label{Ry244u}
R_2^{(3)}={t_{32}}\log_{2}\left(1+\rho\frac{\text{tr}\left(\boldsymbol \psi_{1}\mathbf W_{41}'\right)}{t_{32}}+\rho\frac{\text{tr}\left(\boldsymbol \psi_{2}\mathbf W_{42}\right)}{t_{32}}\right).
\end{equation}
Meanwhile, the energy constraint in (\ref{con}) can be reformed as
\begin{equation}
\label{ncon2}
\tau_{2j}+\tau_{3j}+{\tau_{3j}'}\le \eta P_1\text{tr}\left(\boldsymbol \psi_{j}\mathbf W_{1}\right), i=1,2.
\end{equation}

 Notice that the achievable data rate $R_1^{(2)}$ in (\ref{Ry11u}) is a concave function in $(\mathbf W_2, t_{21})$, and similarly for the rate expressions in (\ref{Ry021u})-(\ref{Ry244u}) (see the proof in \cite{2019:Zheng}). Nonetheless, $\mathbf W_1$ needs to satisfy the non-convex constraint $\text{rank}(\mathbf W_1)=1$, and so do $\mathbf W_{2},\mathbf W_{3},\mathbf W_{41},\mathbf W_{41}',\mathbf W_{42}$ and $\mathbf W_{42}'$. We denote $\boldsymbol\tau=[\tau_{21},\tau_{22},\tau_{31},\tau_{31}',\tau_{32},\tau_{32}']$ and $\widetilde{\mathbf W}=[\mathbf W_1,\mathbf W_{2},\mathbf W_{3},\mathbf W_{41},\mathbf W_{41}',\mathbf W_{42},\mathbf W_{42}']$. We first drop the non-convex rank-one constraints and relax (P2) into the following problem,
\begin{equation}
\begin{aligned}
\label{3}
    ~(\rm{P3}):&~\max_{\overline{R},\mathbf{t},\boldsymbol{\tau},\widetilde{\mathbf W}} & &  ~\overline{R}\\
    &~~~~\text{s. t.} && t_1,t_{2j},t_{3j},\tau_{2j},\tau_{3j},\tau_{3j}'\ge0, j=1,2,\\
    & & & (\ref{t}),(\ref{u}) \ \text{and}\ (\ref{ncon2}),~~~~~\\
    & & & \overline{R}\leq R_1^{(2)},\ \overline{R}\leq R_0^{(21)}+R_1^{(3)},\\
    & & & \overline{R}\leq R_2^{(2)}, \ \overline{R}\leq R_0^{(22)}+R_2^{(3)},\\
    & & & \mathbf W_i, \mathbf W_{4j}, \mathbf W_{4j}'\succeq 0, i=1,2,3,j=1,2.
\end{aligned}
\end{equation}

(P3) is a standard semidefinite programming  (SDP), which can be efficiently solved by convex tools such as CVX \cite{2004:Boyd}. Let us denote the optimal solution to (P3) as $\{\overline{R}^*,\mathbf{t}^*,\boldsymbol{\tau}^*,\widetilde{\mathbf W}^*\}$, we can obtain the optimal $\mathbf V_1^*=\mathbf W_1^*/t_1^*$, $\mathbf V_2^*=\mathbf W_{2}^*/\tau_{21}^*$, $\mathbf V_3^*=\mathbf W_{3}^*/\tau_{22}^*$, $\mathbf V_4^*=\mathbf W_{41}^*/\tau_{31}^*$, $P_{2j}^*=\tau_{2j}^*/t_{2j}^*, P_{3j}^*=\tau_{3j}^*/t_{3j}^*, j=1,2$. However, the relaxed problem (P3) may not lead to a rank-one solution in general. Then, the Gaussian randomization method is employed to construct a rank-one solution. Specifically, to recover $\bar{\boldsymbol v}_1$ from $\mathbf V_1^*$, we obtain the eigenvalue decomposition of $\mathbf V_1^*$ as $\mathbf V_1^*={\boldsymbol U}{\boldsymbol\Sigma} {\boldsymbol U}^H$ \cite{2018:WQQ2}, where ${\boldsymbol U}\in \mathbb {C}^{(N+1)\times (N+1)}$ and ${\boldsymbol \Sigma}\in \mathbb {C}^{(N+1)\times (N+1)}$ denote a unitary matrix and diagonal matrix, respectively. Then, we denote $\bar{\boldsymbol v}_1={\boldsymbol U}{\boldsymbol\Sigma}^{1/2}{\boldsymbol r}$ as a suboptimal solution, where ${\boldsymbol r}\in \mathbb {C}^{(N+1)\times 1}$ is a random vector generated according to ${\boldsymbol r} \sim \mathcal {CN}(\boldsymbol 0 ,\boldsymbol I_{N+1})$. With independently generated Gaussian random vector, we select the optimal $\bar{\boldsymbol v}_1^*$ among all $\boldsymbol r$ to achieve the maximum objective function value of (P3). Finally, we obtain $\boldsymbol v_1^*=e^{j\arg([\bar{\boldsymbol v}_1^*]_{(1:N)}/\bar {v}_{1,N+1}^*)}$, where $\arg(\cdot)$ denotes the phase extraction operation and $[\boldsymbol a]_{(1:N)}$ denotes the vector that contains the first $N$ elements of $\boldsymbol a$. The optimal $\boldsymbol \Theta_1^*$ can be obtained from $\boldsymbol v_1^*$. Following the similar procedure, we can recover $\boldsymbol v_i^*,i=2,3,4$ from $\mathbf V_i^*$, and further obtain the optimal $\boldsymbol \Theta_i^*$ from $\boldsymbol v_i^*$.
\section{Simulation Results}
 \begin{figure}
  \centering
   \begin{center}
      \includegraphics[width=0.46\textwidth]{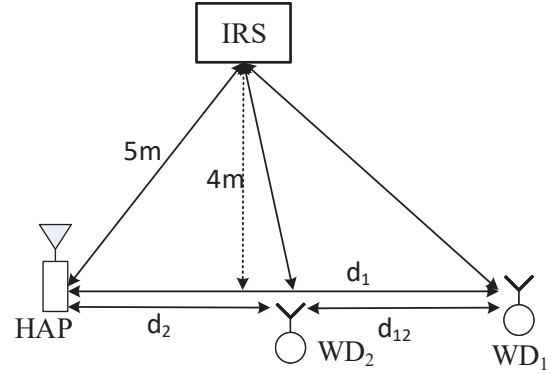}
   \end{center}
  \caption{The placement model of simulation setup.}
  \label{Fig.2}
\end{figure}
In this section, we provide simulation results to evaluate the performance of the proposed IRS-assisted cooperation scheme. To account for small-scale
fading, we assume that all channels follow Rayleigh fading and the distance-dependent path loss is modeled as $L=C_0(\frac{d_i}{d_0})^{-\lambda}$, where $C_0=30$ dB is the path loss at the reference distance $d_0=1$ m, $d_i,i=1,2$, and $d_{12}$ denote the HAP-WD$_i$ and WD$_1$-WD$_2$ distance, and $\lambda$ denotes the path loss exponent. To account for heterogeneous channel conditions, we set different path loss exponents of the HAP-IRS, IRS-WD$_i$, HAP-WD$_i$, WD$_1$-WD$_2$ channels as $2.0, 2.2, 3.0, 3.0$, respectively. Other required parameters are set as $P_1=30$ dBm, $\eta=0.8$, and $N_0=- 80$ dBm.  All the simulation results are obtained by averaging over 1000 channel realizations. For performance comparison, we consider the following representative benchmark methods:
\begin{enumerate}
  \item Independent transmission with IRS: This method follows the harvest-then-transmit protocol in \cite{2014:Ju1}. Specifically, IRS is used to reflect RF energy from the HAP in the FL and WDs's information in the RL.
  \item Information exchange without IRS: This corresponds to the two-user cooperation method in \cite{2017:Zhong}. The detailed expressions are omitted here due to the page limit.
  \item Independent transmission without IRS: WDs transmit their information independently in a round-robin manner to the HAP.
\end{enumerate}

For fair comparison, we optimize the resource allocations in all the benchmark schemes, where the details are omitted due to the page limit.

\begin{figure}
  \centering
   \begin{center}
      \includegraphics[width=0.48\textwidth]{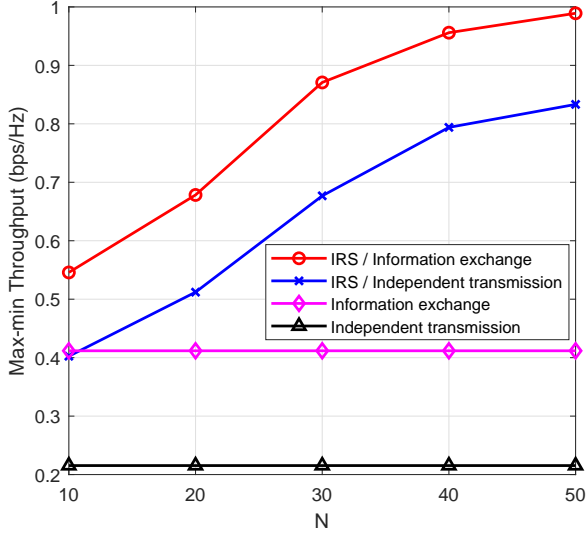}
   \end{center}
  \caption{The max-min throughput performance versus the number of reflecting elements $N$.}
  \label{Fig.3}
\end{figure}
\begin{figure}
  \centering
   \begin{center}
      \includegraphics[width=0.48\textwidth]{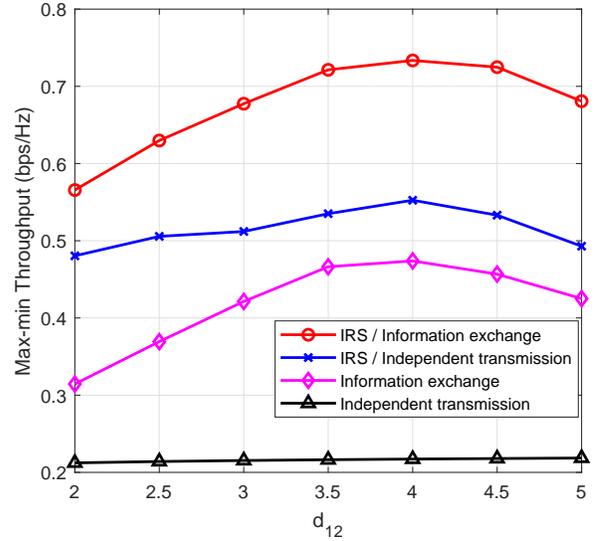}
   \end{center}
  \caption{The max-min throughput performance versus the inter-user channel.}
  \label{Fig.4}
\end{figure}

We consider the placement model of the network system in Fig.~\ref{Fig.2}. Fig.~\ref{Fig.3} first shows the impact of the numbers of reflecting elements $N$ to the throughput performance. Without loss of generality, we set $d_1 = 8$ m, $d_2=5$ m and $d_{12} = 3$ m as a constant and change the value of $N$ from 10 to 50.  Obviously, the two IRS-assisted transmission methods achieve higher throughput due to the array gain.
On average, the proposed IRS-assisted cooperation method achieves $30.17\%$, $102.23\%$ and $275.11\%$ higher throughput than the three benchmark methods, respectively.

Fig.~\ref{Fig.4} investigates the throughput performance versus the inter-user channel $h_{12}$. Here, we still use the placement model in Fig. 2, where we set $d_1=8$ m, $N=20$ and vary $d_{12}$ from 2 to 5 meters. %Thus, the IRS-WD$_2$ link distance is given by $d_{i2}=\sqrt{(d_2-3)^2+4^2}$ m.
Notice that the IRS-assisted communication methods always produce better performance than the other methods without IRS.
We observe that the throughput of the independent transmission method is almost unchanged when $d_{12}$ increases, because the performance bottleneck is the weak channel $h_1$ of the far user WD$_1$.  It is observed that the throughput of the other three methods first increase when $d_{12}\textless 4$ m, but decrease as $d_{12}$ further increases. This is because when WD$_2$ moves closer from the HAP and IRS, the signal gleaned from both the HAP and IRS become stronger. However, as we further increase $d_{12}$, the weak inter-user channel results in the degradation of the communication performance. Besides, the performance gap between the two IRS-assisted methods gradually increases with $d_{12}$. This shows that a weaker inter-user channel (larger $d_{12}$) leads to less efficient cooperation between the two users. Nonetheless, there exists significant performance gap between the two cooperation methods either with or without the use of IRS, indicating the effective performance enhancement of IRS in both energy and information transmissions.
\begin{figure}
  \centering
   \begin{center}
      \includegraphics[width=0.48\textwidth]{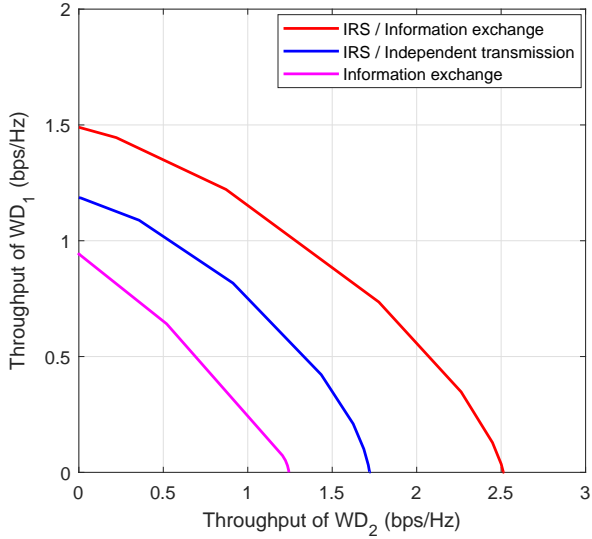}
   \end{center}
  \caption{The achievable rate region comparison of three different methods.}
  \label{Fig.5}
\end{figure}

Fig.~\ref{Fig.5} compares the achievable rate regions of three different schemes, i.e., the proposed IRS-assisted information exchange and independent transmission either with or without the assist of IRS. The rate region can be obtained by replacing the objective of problem (P1) with the weighted sum rate of the two users, i.e., $\omega R_1+ (1 -\omega)R_2$, and solve the optimization under different weighting factor $\omega$ from 0 to 1. The details are omitted due to the page limit. Similarly, we use the placement model in Fig.~\ref{Fig.2} with fixed $d_1 = 8$ m, $d_2 = 5$ m, $d_{12} = 3$ m and $N = 20$. Evidently, we see that the rate region of the proposed IRS-assisted cooperation method is significantly larger than that of the other two methods. On average, it achieves $25.59\%$ and $57.98\%$ higher throughput for WD$_1$,  $45.99\%$ and $102.04\%$ higher throughput for WD$_2$ than the two benchmark methods, respectively. This indicates that the two users can benefit significantly both from the proposed cooperation and the use of IRS.

 The simulation results in Fig.~\ref{Fig.3}, Fig.~\ref{Fig.4} and Fig.~\ref{Fig.5} demonstrate the advantage of applying the IRS to enhance the throughout performance both users when cooperation is considered in WPCN. Besides, the effective enhancement of energy efficiency and spectrum efficiency can benefit from the utilization of IRS.

\section{Conclusions}
In this paper, we investigated the use of IRS in assisting the transmissions in a two-user cooperative WPCN. We formulated an optimization problem to maximize the common throughput. An efficient algorithm is proposed to jointly optimize the phase shifts of the IRS on reflecting the wireless energy and information
signals, the transmission time and power allocation of the two WDs on wireless energy and information transmissions.  Extensive simulations verified that the proposed IRS-assisted user cooperation method can effectively improve the throughput performance in WPCNs under different practical network setups.

\end{document}